\documentclass[aps,prc,twocolumn,floatfix,nofootinbib,
superscriptaddress,amsmath,amssymb,longbibliography]{revtex4-1}

\usepackage{graphicx}
\usepackage{multirow}
\usepackage{amsmath,amssymb}
\usepackage[dvipsnames]{xcolor}
\usepackage[colorlinks=true,linkcolor=blue,citecolor=blue,urlcolor=blue]{hyperref}

\newcommand{\bb}[1]{#1}

\begin{document}

\title{Constrained coupled-cluster computations yield negative-parity rotational bands}

\author{G.~Hagen}
\affiliation{Physics Division, Oak Ridge National Laboratory, Oak Ridge,
Tennessee 37831, USA}
\affiliation{Department of Physics and Astronomy, University of Tennessee,
Knoxville, Tennessee 37996, USA}
\affiliation{Department of Physics and Astronomy, Chalmers University of Technology,
SE-412 96 G\"oteborg, Sweden}

\author{B.~Bally}
\affiliation{Technische Universit\"at Darmstadt, Department of Physics, 64289 Darmstadt, Germany}

\author{T.~Papenbrock}
\affiliation{Department of Physics and Astronomy, University of Tennessee,
Knoxville, Tennessee 37996, USA}
\affiliation{Physics Division, Oak Ridge National Laboratory, Oak Ridge,
Tennessee 37831, USA}

\begin{abstract}
  Many deformed nuclei exhibit low-lying rotational bands with
  negative parity.  In mean-field calculations, such bands follow from
  symmetry restoration when the variational solution spontaneously
  breaks parity.  However, they can also be computed in cases where
  the breaking of parity is not spontaneous but rather enforced by an
  octupole constraint. In this paper, we extend these techniques to
  coupled-cluster computations. We introduce octupole-constrained
  coupled-cluster theory, which explicitly breaks the parity of the
  correlated intrinsic state and restores parity and angular momentum
  as good quantum numbers by projection after variation.  We perform
  coupled-cluster singles-and-doubles (CCSD) computations for
  $^{20}$Ne using \bb{two-} and three-nucleon forces from chiral
  effective field theory.  The energies of the positive-parity
  ground-state rotational band are close to data and depend only
  weakly on the constrained octupole moment and the Hartree--Fock
  reference. In contrast, the energies of the negative-parity
  rotational band are more sensitive to these auxiliary quantities.
\end{abstract}

\maketitle

\section{Introduction} 
Atomic nuclei are \bb{complex} finite quantum systems and typically
exhibit single-particle and collective degrees of freedom \bb{that
  manifest themselves at a similar energy
  scale}~\cite{bohr1969,bohr1975}.  The former can often be understood
\bb{intuitively} within the \bb{traditional} shell model \bb{picture},
while the latter are \bb{more} naturally described in terms of
spontaneous (or emergent) symmetry breaking~\cite{yannouleas2007}.  In
a finite nucleus, \bb{however}, the exact eigenstates retain the
symmetries of the Hamiltonian. \bb{The notion of} ``symmetry
breaking'' refers here to an \bb{intermediate reference} mean-field or
truncated many-body state that breaks a symmetry of the underlying
Hamiltonian.  Symmetry projection of \bb{this so-called}
\textit{intrinsic} state then yields states with good quantum numbers
and the associated collective spectrum~\cite{lipkin1960,yao2020,
  sheikh2021,bally2021, frosini2022b, hagen2022,
  sun2025,zhou2025a,zhou2026}.  Examples include ground-state
rotational bands in open-shell deformed nuclei and pairing rotations
in superfluid nuclei~\cite{bohr1975,brink2005,coelloperez2025}. This
framework also describes pear-shaped nuclei whose intrinsic states
spontaneously break reflection
symmetry~\cite{nazarewicz1984,egido1991,butler1996}.  Such states are
of renewed interest because octupole collectivity can enhance nuclear
Schiff moments relevant to searches for a permanent electric dipole
moment~\cite{auerbach1996,spevak1997,engel2013}.

In many nuclei, however, low-lying bands of opposite parity exist even
though the mean-field solution does not break parity spontaneously.
The nucleus $^{20}$Ne is one such
example~\cite{zhou2016,marevic2018,frosini2022b}: its Hartree--Fock
minimum energy exhibits quadrupole deformation, axial symmetry, and
reflection symmetry. To obtain negative-parity states from
symmetry-projected intrinsic states, one has to break reflection
symmetry explicitly, e.g., by an octupole constraint in the mean-field
calculation.  Projection of the resulting intrinsic states onto good
angular momentum and parity then yields low-lying states with good
spin and parity quantum numbers.

This mean-field strategy also applies to single-reference \textit{ab
  initio} methods~\cite{hergert2020,ekstrom2023}, where correlations
are built on top of a symmetry-broken reference state.  Examples are
coupled-cluster theory, the in-medium similarity renormalization
group, self-consistent Green's function theory, and many-body
perturbation
theory~\cite{dickhoff2004,hagen2014,hergert2016,soma2020b,tichai2020}.
Typically these methods start from a microscopic Hamiltonian from
chiral effective field
theory~\cite{weinberg1990,epelbaum2009,machleidt2011,Hammer:2019poc}
and contain \bb{two-} and three-nucleon forces. As for mean-field
methods, breaking a symmetry in the reference state is now a standard
route to open-shell nuclei.  Breaking a global $U(1)$ symmetry
underlies Gorkov Green's function and Bogoliubov coupled-cluster (BCC)
calculations~\cite{soma2013,soma2020b,soma2021,signoracci2015,tichai2024},
whereas breaking rotational symmetry has enabled deformed
coupled-cluster and related expansion
methods~\cite{duguet2015,hagen2022,sun2025}.  Most recently, deformed
self-consistent Green's function theory has been implemented through
third order~\cite{scalesi2026a} and applied to the heavy, doubly
open-shell deformed nucleus $^{238}$U~\cite{scalesi2026b}.

Symmetry restoration is the final step in this approach.  Rotational
symmetry restoration in coupled-cluster theory was formulated
by~\textcite{duguet2015} and subsequently realized with disentangled
projected coupled-cluster methods~\cite{qiu2017,hagen2022,sun2025}.
In the superfluid case, global gauge restoration and particle-number
projection were formulated for BCC~\cite{duguet2017a} and tested in a
particle-number-projected BCC implementation~\cite{qiu2019}.  The
combined angular-momentum and parity projection employed below builds
on these developments.

Adding an octupole constraint to the Hamiltonian via a Lagrange
multiplier generates states with broken parity.  In such an approach,
physical energies are evaluated using the Hamiltonian with zero
Lagrange multiplier. Consider an octupole-deformed product state that
serves as the reference for the coupled-cluster computation, and
suppose that parity is not spontaneously broken, i.e., the
octupole-constrained reference state has a higher energy than the
optimal mean-field reference (which then has good parity). Then, the
coupled-cluster singles amplitudes tend to relax the orbitals toward
the lower-energy, reflection-symmetric Hartree--Fock minimum. One
might think that discarding those excitations, i.e., using coupled
cluster with doubles instead of coupled cluster with singles and
doubles, might avoid this problem.  However, such an approach is too
crude because it also removes the orbital relaxation required in the
presence of two-particle--two-hole correlations. A way forward is to
also include the octupole constraint in the correlated calculation.
Furthermore, because the laboratory-coordinate octupole operator is
origin dependent, one also needs to enforce an additional dipole
constraint to keep the center-of-mass of the correlated many-body
state at the origin. One can then, as the last step, perform symmetry
projection and obtain states with definite angular momentum and
parity.

This paper introduces constrained coupled-cluster theory as a method
to access collective states that result from explicit (rather than
spontaneous) symmetry breaking. We note that the case of spontaneous
symmetry breaking is trivially included in our approach as it requires
no octupole constraint in coupled-cluster computations.  The enforced
breaking of reflection symmetry yields correlated intrinsic states,
which -- after symmetry restoration -- give us access to
negative-parity bands even when such an asymmetry does not emerge from
unconstrained Hartree--Fock computations.

This paper is organized as follows. We describe the Hamiltonian and
employed model space in Sec.~\ref{sec:ham}. The 
constrained coupled-cluster approach is presented in
Sec.~\ref{sec:constrain}. The ensuing symmetry projection is
described in Sec.~\ref{sec:proj}. We apply this approach to $^{20}$Ne
and compute its positive-parity ground-state rotational band and the
negative-parity band that is associated with breaking of parity; the
results are presented in Sec.~\ref{sec:results}. Finally, we present
a summary in Sec.~\ref{sec:sum}.

\section{Hamiltonian and model space}
\label{sec:ham}

We start from the intrinsic Hamiltonian
\begin{equation}
 H = T-T_{\mathrm{CM}}+V_{\mathrm{2N}}+V_{\mathrm{3N}},
\end{equation}
where the kinetic energy of the center of mass, $T_\mathrm{CM}$, is
removed from the total kinetic energy $T$, and employ the
$1.8/2.0$~(EM) interaction~\cite{hebeler2011} for the \bb{two-} and
three-nucleon potentials $V_{2N}$ and $V_{3N}$, respectively.  The
Hamiltonian is represented in a single-particle basis of
\bb{spherical} harmonic-oscillator states with radial quantum number
$n$ and orbital angular momentum $\ell$. We include states with up to
$2n+\ell\leq e_{\max}=8$ quanta of excitation, and use an oscillator
spacing of $\hbar\omega=16$~MeV.  For the three-nucleon interaction we
apply the additional energy cut $E_{3,\max} = 16\hbar\omega$ on the
matrix elements. We use the two-body rank reduction of
Ref.~\cite{frosini2021}; normal ordering with respect to a spherical
Hartree--Fock reference preserves rotational invariance of the
resulting two-body Hamiltonian.  We perform axially symmetric
Hartree--Fock computations with an octupole constraint and a dipole
constraint. This yields the single-particle basis and the reference
state $|\Phi_0\rangle$.  In what follows, we present coupled-cluster
results starting from Hartree--Fock references that were constrained
to mass octupole moments $Q_{30}^{\rm HF}=75$, $100$, $125$, and
$150$~fm$^3$ \bb{(see below for the definition of the multipole
  operators)}.

We compare these results with projected Hartree--Fock (PHF)
\bb{calculations performed with the same chiral Hamiltonian.  More
  precisely, we systematically explore the angular-momentum ($J$) and
  parity ($\pi$) projected energy surface as a function of the
  constraints $(\langle Q_{20}\rangle, \langle Q_{30} \rangle)$ on the
  underlying mean-field reference states and determine the minimum for
  each value of $J^\pi$. We first perform a broad scan of the energy
  surface and then, in the region of the minima, use a finer mesh for
  the constraints with steps
  $\Delta \langle Q_{20}\rangle \approx 2.5~\text{fm}^2$ and
  $\Delta \langle Q_{30}\rangle \approx 8.25~\text{fm}^3$. The PHF
  calculations were performed with the numerical suite
  TAURUS~\cite{bally2021b,bally2024a}.}

\section{Constrained coupled-cluster theory}
\label{sec:constrain}

Coupled-cluster theory has been reviewed in
Refs.~\cite{kuemmel1978,bishop1991,bartlett2007,hagen2014} and the
textbook~\cite{shavittbartlett2009}. Here, we only present a few key
concepts of this well-known method.  We use the coupled-cluster
singles-and-doubles (CCSD) approximation, where the right eigenstate
is \bb{expanded as}
\begin{equation}
    |\Psi\rangle=e^T|\Phi_0\rangle \ , 
\end{equation}
where $T=T_1+T_2$ consists of one-particle--one-hole and
two-particle--two-hole excitations. The left eigenstate is
\bb{expressed as}
\begin{equation}
   \langle\widetilde\Psi|=\langle\Phi_0|(1+\Lambda)e^{-T} \ , 
\end{equation}
where $\Lambda=\Lambda_1+\Lambda_2$ \bb{is a de-excitation operator
  containing one- and two-body components}. The constrained
Hamiltonian is
\begin{equation}
 H'(\lambda_1,\lambda_3)=H-\lambda_1Q_{10}-\lambda_3Q_{30}.
 \label{eq:constrained-hamiltonian}
\end{equation}
Here, $\lambda_1$ and $\lambda_3$ are Lagrange multipliers and
$Q_{\mu 0}=\sum_{i=1}^A r_i^\mu Y_{\mu 0}(\widehat{\boldsymbol r}_i)$
with $\mu=1,3$ are mass multipole operators; hence $Q_{10}$ and
$Q_{30}$ have units fm and fm$^3$, respectively.  These
laboratory-coordinate operators are origin dependent; in particular, a
translation of the reference state mixes its dipole, quadrupole, and
octupole moments. This is the reason we employ the dipole constraint
\bb{to fix, on average, the position of the center of mass}.
Following the terminology of constrained mean-field
theory~\cite{ring1980}, we refer to $H'$ as the Routhian.

Introducing the similarity transformed operator $\overline O=e^{-T}Oe^T$,
the bivariational energy functional becomes
\begin{equation}
 {\cal E}'(T,\Lambda,\boldsymbol\lambda)
 =\langle\Phi_0|(1+\Lambda)\overline {H'}(\boldsymbol\lambda)|\Phi_0\rangle.
 \label{eq:bivariational-functional}
\end{equation}
Here we introduced $\boldsymbol\lambda=(\lambda_1,\lambda_3)$ to
condense the notation of the Lagrange multipliers.  Stationarity with
respect to the independent left and right amplitudes gives
\begin{equation}
 \frac{\partial {\cal E}'}{\partial T}=0,
 \qquad
 \frac{\partial {\cal E}'}{\partial\Lambda}=0.
 \label{eq:stationarity}
\end{equation}
The multipole moments of the resulting state are
\begin{equation}
 \langle Q_{\mu 0}\rangle
 =\langle\Phi_0|(1+\Lambda)\overline Q_{\mu 0}|\Phi_0\rangle .
 \label{eq:constraint-expectation}
\end{equation}
Here $\langle Q_{\mu 0}\rangle$ is the moment of the
\emph{unprojected} biorthogonal CCSD state.  The octupole moment
$\langle Q_{30}\rangle$ labels the constrained intrinsic state,
whereas $\langle Q_{10}\rangle$ fixes the center of mass; neither
should be interpreted as a laboratory moment of the state obtained
after $J^\pi$ projection.  In mean-field calculations, constrained
solutions with different $\langle Q_{30}\rangle$ are commonly used as
generator states in the projected generator coordinate method
(PGCM)~\cite{hill1953,griffin1957,robledo2011a,bernard2016a,frosini2022b,giacalone2025a};
we explore the analogous mixing of constrained coupled-cluster states
in Sec.~\ref{sec:results}.  Variation with respect to $\Lambda$
produces the usual nonlinear right CCSD equations for the Routhian,
\begin{align}
 E'(\boldsymbol\lambda)&=\langle\Phi_0|\overline {H'}|\Phi_0\rangle,\nonumber\\
 0&=\langle\Phi_i^a|\overline {H'}|\Phi_0\rangle,\nonumber\\
 0&=\langle\Phi_{ij}^{ab}|\overline{H'}|\Phi_0\rangle,
 \label{eq:constrained-ccsd}
\end{align}
while variation with respect to $T$ gives the linear left-amplitude
equations.  Thus both $T$ and $\Lambda$ must be recomputed whenever the
Lagrange multipliers change.  
%Once $T$ and $\Lambda$ 
Once these are known, one can compute the energy $E'(\boldsymbol\lambda)$
and, via Eq.~\eqref{eq:physical-energy} below, the physical energy $E$.

At each value of $\lambda_3$ we adjust $\lambda_1$ such that
\begin{equation}
 \langle Q_{10}\rangle=0.
 \label{eq:dipole-constraint}
\end{equation}
An exact eigenstate of the intrinsic Hamiltonian factorizes into an
intrinsic and a center-of-mass part, and coupled-cluster wave
functions exhibit this factorization to a good
approximation~\cite{hagen2009a}.  The octupole constraint breaks
translational invariance, however, because $Q_{30}$ is defined in
laboratory coordinates, and a displacement of the density then mimics
an octupole moment~\cite{ring1980}. Thus it is imperative to keep the
center of mass centered at the origin via the dipole constraint.  We
note that imposing this constraint via
Eq.~\eqref{eq:dipole-constraint} is also standard in multireference
density-functional calculations of octupole
states~\cite{robledo2011a,yao2015,yao2016,zhou2025b}.  Without the
dipole constraint, CCSD calculations starting from octupole-deformed
Hartree--Fock references with $\langle Q_{10}\rangle=0$ develop dipole
moments as large as $1.4$~fm.  Using only the dipole constraint in
CCSD computations collapses the octupole moment of the reference
state, $Q_{30}^{\rm HF}$, from about $100$~fm$^3$ to
$\langle Q_{30}\rangle<20$~fm$^3$, because $e^{T_1}$ is a Thouless
transformation that rotates the orbitals back toward the
parity-conserving Hartree--Fock minimum.  We therefore keep the dipole
constraint and vary $\lambda_3$ to generate CCSD states over the range
of octupole moments needed for the projection.  No quadrupole
multiplier is applied, and $\langle Q_{20}\rangle$ relaxes
self-consistently to a nonzero value.  In practice, we adjust
$\lambda_1$ by a secant iteration on
$\langle Q_{10}\rangle(\lambda_1)$, reconverging both the right and
left CCSD amplitudes at every step, and retain only solutions with
$|\langle Q_{10}\rangle|\leq0.005$~fm.

Once we obtain converged CCSD results for a given constraint, the physical
energy is evaluated with the unconstrained Hamiltonian,
\begin{align}
 E&=\langle\Phi_0|(1+\Lambda)e^{-T}He^T|\Phi_0\rangle\nonumber\\
  &=E'+\lambda_1\langle Q_{10}\rangle+\lambda_3\langle Q_{30}\rangle
  =E'+\lambda_3\langle Q_{30}\rangle,
 \label{eq:physical-energy}
\end{align}
rather than with the Routhian $H'$; the last equality uses
Eq.~\eqref{eq:dipole-constraint}.  The constrained state is an
approximate eigenstate of $H'$, not of $H$.  Symmetry projection
yields good quantum numbers for parity and angular momentum. However,
projection after variation does not yield an exact eigenstate of $H$
because CCSD is not exact and neither is the truncation of the
disentangled expansion used in the symmetry projection. Thus, the
projected energy can exhibit a dependence on both the constraint and
the generating Hartree--Fock reference; we use this dependence as a
diagnostic below.

\bb{Finally, we mention that a similar constrained scheme has been
  employed within BCC theory to fix the average number of
  particles~\cite{demol2024a,demol2026a,vernik2026a}.}

\section{Symmetry projection} 
\label{sec:proj}

Parity projection of octupole-deformed intrinsic states has a long
history in mean-field computations~\cite{egido1991}.  Here we project
each constrained CCSD solution with the disentangled coupled-cluster
formulation of~\textcite{qiu2017}. The results reported below use the
angular-momentum- and parity-projected coupled-cluster bivariational
functional~\cite{qiu2017,sun2025},
\begin{equation}
 E(J^\pi)=
 \frac{\langle\widetilde\Psi|P^{J\pi}H|\Psi\rangle}
 {\langle\widetilde\Psi|P^{J\pi}|\Psi\rangle} \ ,
 \label{eq:projected-lr-functional}
\end{equation} 
where the projection operator
\begin{equation}
    P^{J\pi}=P^JP^\pi 
\end{equation}
is a product of angular-momentum and parity projection operators. 

Thus both the right state $e^T|\Phi_0\rangle$ and the left state
$\langle\Phi_0|(1+\Lambda)e^{-T}$ enter the projected energy; this is
the projected linear-response CCSD approach and we refer to this
simply as projected CCSD, abbreviated as PCCSD.  Numerically,
Eq.~\eqref{eq:projected-lr-functional} is assembled from the
angle-dependent norm and Hamiltonian kernels
${\cal N}=\langle\widetilde\Psi|R(\beta,\Theta)|\Psi\rangle$ and
${\cal H}=\langle\widetilde\Psi| R(\beta,\Theta) H|\Psi\rangle$, see
Refs.~\cite{qiu2017,sun2025} for details.

For an axially symmetric reference state with \bb{a vanishing
  angular-momentum projection along the $z$-axis, $K=0$}, the
projectors \bb{read}
\begin{align}
 P^J&=\frac{2J+1}{2}\int_0^\pi d\beta\,\sin\beta\,
 d^J_{00}(\beta)R(\beta),\nonumber\\
 P^\pi&=\sum_{\Theta=0,\pi}\omega_\Theta^\pi e^{i\Theta\hat N_-},
 \qquad(\omega_0^{\pm},\omega_\pi^{\pm})= (1/2,\pm1/2) \ .
 \label{eq:parity-projector}
\end{align}
Here $\hat N_-=\sum_{k}a_k^\dagger a_k$ counts the particles in
negative-parity orbitals (the sum runs over those orbitals only), and
the \bb{label} $\pi=\pm$ denotes the parity of the projected
state. Finally, $R(\beta)=e^{-i\beta J_y}$ is a rotation operator and
$d^J_{00}(\beta)$ the small Wigner $d$ matrix. Expressing the parity
projection via an exponential -- as is done for the angular-momentum
projection -- is essential for our approach.

The projection employs the unitary transformation
$R(\beta,\Theta)=R(\beta)e^{i\Theta\hat N_-}$.  The modified Thouless theorem
then gives
\begin{equation}
 \langle\Phi_0|R(\beta,\Theta)
 =n(\beta,\Theta)\langle\Phi_0|e^{V(\beta,\Theta)},
 \label{eq:modified-thouless}
\end{equation}
where $V$ is a one-particle--one-hole de-excitation operator and $n$ is the
determinant overlap.
The rotated correlated state is then computed in the disentangled formalism as 
\begin{equation}
 e^{V(\beta,\Theta)}e^T|\Phi_0\rangle
 =e^{W_0(\beta,\Theta)+W_1(\beta,\Theta)+W_2(\beta,\Theta)+\cdots}|\Phi_0\rangle.
 \label{eq:parity-disentangling}
\end{equation}
Here $W_k$ is a $k$-particle--$k$-hole excitation operator.
Parity projection therefore uses the same
angle-dependent differential equations as angular-momentum
projection~\cite{qiu2017,sun2025} with the addition of the two
discrete values $\Theta=0,\pi$ for parity, and their kernels are
combined with the weights in Eq.~\eqref{eq:parity-projector}.  

In this work we truncate the expansion in
Eq.~\eqref{eq:parity-disentangling} at $W_0+W_1+W_2$, and the
amplitudes are obtained by solving a set of ordinary differential
equations~\cite{qiu2017,sun2025}. Note that even though we truncate
$T=T_1 + T_2$, rotating the unprojected state induces terms up to
$W_A$; angular-momentum and parity projection are therefore
approximate rather than exact. In particular, an exact
angular-momentum-projected state satisfies
$\langle J^2\rangle=J(J+1)$, so deviations from this value directly
measure the residual error associated with the truncation at $W_2$.
The $\beta$ integrals are evaluated with a 20-point Gauss--Legendre
mesh on $[0,\pi]$.  We checked the convergence of the angular
quadrature by increasing the mesh size; the low-$J$ energies presented
below are stable at the displayed precision.

\begin{figure*}[t]
 \centering
 \includegraphics[width=0.90\textwidth]{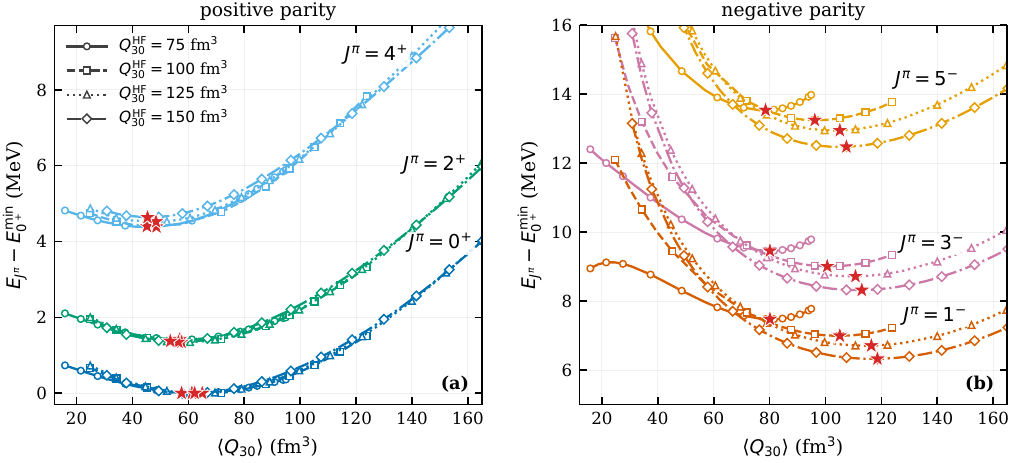}
 \caption{Projected CCSD excitation energies of the (a)
   positive-parity and (b) negative-parity states as functions of the
   octupole moment $\langle Q_{30}\rangle$ of the unprojected CCSD
   state, for the four Hartree--Fock references with
   $Q_{30}^{\rm HF}=75$~fm$^3$ (solid lines with open circles),
   $100$~fm$^3$ (dashed lines with open squares), $125$~fm$^3$ (dotted
   lines with open triangles), and $150$~fm$^3$ (dash-dotted lines
   with open diamonds). Each energy curve is measured from the minimum
   projected $0^+$ energy of the same reference.  A red star on each
   curve marks its projected-energy minimum.  }
 \label{fig:projected-curves}
\end{figure*}

\section{Results}
\label{sec:results}

Figure~\ref{fig:projected-curves} shows the projected excitation
energies as functions of the octupole moment $\langle Q_{30}\rangle$,
Eq.~\eqref{eq:constraint-expectation}, of the underlying unprojected
CCSD state, starting from four
octupole-constrained Hartree--Fock references with
$Q_{30}^{\rm HF}=75$, $100$, $125$, and $150$~fm$^3$. The unconstrained
Hartree--Fock minimum of $^{20}$Ne preserves parity and has
$\langle Q_{30}\rangle=0$, so the negative-parity projection of that
determinant has zero norm.  For references with small but finite
$Q_{30}^{\rm HF}$ the negative-parity norm is small and the
corresponding kernels are ill-conditioned.  We therefore start the constrained
CCSD trajectories from Hartree--Fock references with sizable
$Q_{30}^{\rm HF}$.

The energies with positive parity from the four different reference
states remain close over the shown range of $\langle
Q_{30}\rangle$. Projection and CCSD correlations remove most of the
memory of the octupole moment $Q_{30}^{\rm HF}$ of the reference
state. The energy minima of the $J^\pi=0^+, 2^+, 4^+$ states occur at
comparatively modest octupole moments: for the
$Q_{30}^{\rm HF}=75$~fm$^3$ reference the $0^+$, $2^+$, and $4^+$
minima are at approximately $61$, $57$, and $49$~fm$^3$, while the
corresponding values from the $Q_{30}^{\rm HF}=100$~fm$^3$ reference
are approximately $58$, $58$, and $45$~fm$^3$.  The resulting
positive-parity excitation energies vary little: $E(2^+)$ is between
$1.35$ and $1.41$~MeV, while $E(4^+)$ changes gradually from $4.39$ to
$4.63$~MeV across the four references.

The energies of the negative-parity states, shown in
Fig.~\ref{fig:projected-curves}(b), behave differently.  Their
energies visibly depend on octupole moment $Q_{30}^{\rm HF}$ of the
reference state, and the spacings are also sensitive to the
constrained value of $\langle Q_{30}\rangle$. \bb{In particular,} the
energy minima of the $J^\pi=1^-, 3^-, 5^-$ states are at higher
\bb{values of} $\langle Q_{30}\rangle$ than for the positive-parity
states, and their positions move from about $80$~fm$^3$ for
$Q_{30}^{\rm HF}=75$~fm$^3$ to about $119$~fm$^3$ for
$Q_{30}^{\rm HF}=150$~fm$^3$ (see Table~\ref{tab:reference-minima}),
whereas the $0^+$ minima stay between $58$ and $65$~fm$^3$.  The
minimum energies of the negative-parity states vary by about $0.5$~MeV
across the four references.  Since the $0^+$ minimum moves in the
opposite direction, the $1^-$ excitation energy measured from the
$0^+$ minimum of the same reference varies by more than $1$~MeV,
compared with less than $0.1$~MeV for the $2^+$ state.

Table~\ref{tab:reference-minima} shows how the projected $0^+$ and
$1^-$ minimum energies depend on the Hartree--Fock reference.  The
positive-parity minimum is most bound for $Q_{30}^{\rm HF}=75$~fm$^3$
and changes by $0.61$~MeV over the four references.  In contrast, the
binding of the $1^-$ state increases with increasing octupole moment,
although the successive gains decrease from $0.37$ to $0.12$ and
$0.06$~MeV.

\begin{table}[t]
  \caption{Dependence of the state-specific PCCSD minima on the
    octupole moment, $Q_{30}^{\rm HF}$ (in fm$^3$), of the
    Hartree--Fock reference.  For each parity, $\langle Q_{30}\rangle$
    (in fm$^3$) is the octupole moment of the underlying constrained
    CCSD state at the projected minimum and $E$ (in MeV) is the
    corresponding absolute projected energy.  In the last column every
    $1^-$ minimum is measured from the globally lowest projected $0^+$
    energy, $E_{0^+}^{\rm global}=-155.677$~MeV, as in
    Fig.~\ref{fig:ne20-bands}.}
\label{tab:reference-minima}
\begin{ruledtabular}
\begin{tabular}{c rc rc c}
\multirow{2}{*}{$Q_{30}^{\rm HF}$} & \multicolumn{2}{c}{$0^+$ minimum} &
\multicolumn{2}{c}{$1^-$ minimum} & \multirow{2}{*}{$E_{1^-}^{\min}-E_{0^+}^{\rm global}$} \\
 & $\langle Q_{30}\rangle$ & $E$ &
$\langle Q_{30}\rangle$  & $E$  &  \\
\hline
75  & 61.27 & $-155.677$ & 80.12  & $-148.200$ & 7.477 \\
100 & 57.56 & $-155.566$ & 105.12 & $-148.572$ & 7.105 \\
125 & 64.97 & $-155.395$ & 116.40 & $-148.687$ & 6.990 \\
150 & 62.24 & $-155.069$ & 118.59 & $-148.744$ & 6.933 \\
\end{tabular}
\end{ruledtabular}
\end{table}

\begin{figure*}[htb]
 \centering
 \includegraphics[width=0.94\textwidth]{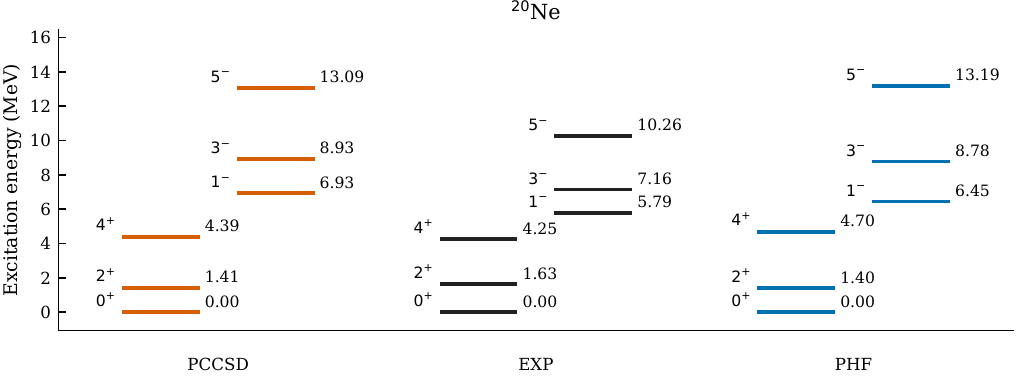}
 \caption{Positive- and negative-parity levels of $^{20}$Ne
 from projected coupled cluster (PCCSD) and projected Hartree--Fock (PHF) at
 $e_{\max}=8$ and $\hbar\omega=16$~MeV, compared with data
 (EXP)~\cite{ensdf2023}.  Each PCCSD level is the state-specific minimum,
 i.e., the lowest projected energy of that $J^\pi$ along the four
 constrained trajectories generated from the $Q_{30}^{\rm HF}=75$, $100$,
 $125$, and $150$~fm$^3$ references; excitation energies are measured from
 the lowest projected $0^+$ energy.  Different levels may therefore come
 from different references and different $\langle Q_{30}\rangle$.  The
 PHF levels are obtained by minimizing the projected energy surface
 separately for each value of $J^\pi$. Positive (negative) parity is shown in the
 left (right) subcolumn, with energies in MeV. 
 \label{fig:ne20-bands}
 }
\end{figure*}

The last column of Table~\ref{tab:reference-minima} is the difference
between minimum $1^-$ energies and the globally lowest $0^+$ energy,
$-155.677$~MeV, obtained from $Q_{30}^{\rm HF}=75$~fm$^3$.  The
resulting excitation decreases from about $7.5$ to about $6.9$~MeV and
changes little beyond $Q_{30}^{\rm HF}=125$~fm$^3$.  For each $J^\pi$
we take the lowest projected energy along the four trajectories, i.e.,
over all $\langle Q_{30}\rangle$ and all four references, and refer to
it as the state-specific minimum.  We measure all excitation energies
from the same $0^+$ energy because the two band heads are minimized by
different Hartree--Fock references. The resulting spectrum is shown in
Fig.~\ref{fig:ne20-bands} \bb{discussed below}.

\begin{figure}[t]
 \centering
 \includegraphics[width=\columnwidth]{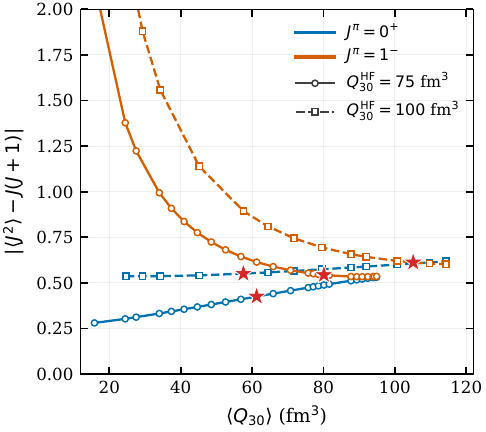}
 \caption{Absolute angular-momentum deviations
   $|\langle J^2\rangle-J(J+1)|$ for the projected $0^+$ and $1^-$
   states at $e_{\max}=8$ and $\hbar\omega=16$~MeV, as functions of
   the octupole moment $\langle Q_{30}\rangle$ of the unprojected CCSD
   state. Solid lines with circles and dashed lines with squares
   denote trajectories generated from Hartree--Fock references
   constrained to $Q_{30}^{\rm HF}=75$ and $100$~fm$^3$,
   respectively. Red stars mark the projected-energy minima of the
   corresponding state and trajectory,
   cf. Fig.~\ref{fig:projected-curves}.}
 \label{fig:j2-quality}
\end{figure}

As a diagnostic for the accuracy of the angular-momentum projection in
the disentangled coupled-cluster approach, we computed the absolute
deviation of $\langle J^2\rangle$ from $J(J+1)$.
Figure~\ref{fig:j2-quality} shows this diagnostic for the $0^+$ and
$1^-$ band heads in constrained
coupled-cluster computations starting from the two Hartree--Fock
reference states with $Q_{30}^{\rm HF}=75$ and $100$~fm$^3$.  For
the $0^+$ state the deviation is small and changes little along the
constrained paths.  For negative parity it is large at small octupole
moments, where the negative-parity projected norm is small, and it
decreases rapidly with increasing $\langle Q_{30}\rangle$.  It is
smallest, and only moderately larger than for the $0^+$ state,
at and beyond the minima of the projected energies (marked by stars), where the curves in
Fig.~\ref{fig:projected-curves}(b) flatten.

Figure~\ref{fig:ne20-bands} shows the two rotational bands obtained
from \bb{PCCSD calculations with combined angular-momentum and parity
  projections. They are compared to known experimental data for
  $^{20}$Ne as well as theoretical results from PHF
  calculations. Focusing first on the positive parity band built on
  top of the ground state, we see that both computations yield similar
  results, which are in reasonable agreement with experiment. We note,
  however, that the ratio $R_{42}=E(4^+)/E(2^+) \approx 3.11 (3.36)$
  for PCCSD (PHF) calculations lies much closer to the rigid-rotor
  value $10/3$ than the experimental value $2.61$. A similar result
  was also found in Ref.~\cite{sun2025}.  Analyzing then the negative
  parity band built on top of the $1^-$ excited state, we first see
  that the two theoretical schemes provide again a very consistent
  description. But contrary to the positive parity states, in this
  case the calculations yield too large a rotational constant (i.e.,
  too small a moment of inertia) when compared to experiment. Also,
  the band head is too high in energy, by about $1.1$~MeV for PCCSD
  and $0.7$~MeV for PHF.  }

\bb{The good agreement between PCCSD and PHF calculations is
  consistent with previous observations that relative properties such
  as excitation energies are well captured at the projected mean-field
  level \cite{frosini2022b,frosini2022c}. However, to also describe
  bulk properties such as the total binding energy, it is necessary to
  further include dynamical correlations as done here with the CCSD
  expansion.}

\bb{While the discrepancy between theory and experiment may originate
  from the limitations of the nuclear Hamiltonian currently employed,
  it may also have other sources. First, we point out that the
  calculations are not yet fully converged with respect to $e_{\max}$,
  and the results from our $e_{\max}= 8$ calculations may still shift
  as the model space is enlarged.  Furthermore, the results might
  suffer from the shortcomings of the underlying projection after
  variation approach used here. Indeed, optimizing the CCSD amplitudes
  before projection is a plausible contributor. Performing variation
  after projection by varying the projected bivariational
  functional~\cite{qiu2018} could improve the moment of
  inertia. Finally, we remark that a recent large-scale PGCM
  calculation demonstrated the importance of triaxial (quadrupole and
  octupole) deformations in this nucleus~\cite{giacalone2025a}, which
  are omitted in the present study. This is left for future work.  }

\bb{Still, to improve the theoretical description,} we tried to mix
the projected constrained CCSD states of a given Hartree--Fock
reference, i.e., states with different octupole moments along one
trajectory, in the spirit of the PGCM.  \bb{However,} after projection
we find that these states are almost linearly dependent: for a given
$J^\pi$, the second singular value of the normalized norm kernel is
only $1$--$5\%$ of the largest one, and all others are below
$10^{-3}$.  Mixing the two independent combinations changes the
projected energies by less than $0.2$~MeV and does not lower the
state-specific minima shown in Fig.~\ref{fig:ne20-bands}.  A more
general approach that mixes coupled-cluster solutions built on
different, non-orthogonal constrained reference states is also left
for future work.

\section{Summary}
\label{sec:sum}

We developed octupole-constrained CCSD as a tool to compute
negative-parity states when the Hartree--Fock minimum itself preserves
parity. Keeping the octupole constraint active in the CCSD computation
prevents one-particle--one-hole excitations from removing parity
breaking from the correlated state, while an additional dipole
constraint fixes the origin of the laboratory-coordinate octupole
moment. For $^{20}$Ne, projection of the resulting states yields a
positive-parity rotational band close to experiment and a low-lying
negative-parity band computed from chiral two- and three-nucleon
forces. The negative-parity excitation energies are too high, and they
depend more strongly on the octupole constraint and on the
Hartree--Fock reference than the positive-parity energies do.  Larger
model spaces, variation after projection, and the mixing of states
built on different constrained references could reduce these
deficiencies. Constrained coupled-cluster theory thus makes octupole
collective states accessible in nuclei where reflection asymmetry does
not emerge at the mean-field minimum.

\acknowledgments This work was supported by the U.S. Department of
Energy, Office of Science, Office of Nuclear Physics, under Award
No.~DE-FG02-96ER40963, and by SciDAC-5 (NUCLEI
collaboration). G.H. acknowledges support and hospitality from
Chalmers University of Technology through the Chalmers Jubilee
Professorship. This work received funding from the European Research
Council under the European Union's Horizon Europe Research and
Innovation Programme (Grant Agreement No.~101162059). Computer time
was provided by the Innovative and Novel Computational Impact on
Theory and Experiment (INCITE) program.  This research used resources
of the Oak Ridge Leadership Computing Facility at Oak Ridge National
Laboratory, supported by the Office of Science of the Department of
Energy under Contract No.~DE-AC05-00OR22725.  The authors acknowledge
the use of ChatGPT (OpenAI) for assistance with code development and
debugging and for improving the clarity of the manuscript, and of
Claude (Anthropic) for proofreading, rewording, checking quoted
numbers, and figure layout. All scientific results and conclusions
were independently verified by the authors.

\end{document}